\documentclass[12pt,thmsa]{article}
\usepackage{amsfonts}

\usepackage{sw20lart}

\input tcilatex

\begin{document}

\title{\flushleft\textbf{Locality and measurements within the SR model for an objective
interpretation of quantum mechanics}}
\author{\hspace*{-0.1cm}\textbf{Claudio Garola}\thanks{%
Dipartimento di Fisica dell'Universit\`{a} and Sezione INFN, 73100 Lecce,
Italy; e-mail: garola@le.infn.it} \, \textbf{and Jaros\l aw Pykacz}\thanks{%
Instytut Matematyki, Uniwersytet Gda\'{n}ski, 80-952, Gda\'{n}sk, Poland;
e-mail: pykacz@delta.math.univ.gda.pl}}
\date{}
\maketitle

$\,$

\noindent
\hspace*{2.1cm}One of the authors has recently propounded an \textit{SR (semantic \hspace*{2.1cm}realism)
model} which shows, circumventing known no-go theo\-\hspace*{2.1cm}rems, that an objective
(noncontextual, hence local) \textit{\ }interpreta\-\hspace*{2.1cm}tion of quantum mechanics
(QM) is possible. We consider here \hspace*{2.1cm}compound physical systems and show why
the proofs of nonlo\-\hspace*{2.1cm}cality of QM do not hold within the SR model, which is
slightly \hspace*{2.1cm}simplified in this paper. We also discuss quantum measurement
\hspace*{2.1cm}theory within this model, note that the \textit{objectification} problem
\hspace*{2.1cm}disappears since the measurement of any property simply reveals \hspace*{2.1cm}its unknown
value, and show that the projection postulate can be \hspace*{2.1cm}considered as an
approximate law, valid FAPP (for all practical \hspace*{2.1cm}purposes). Finally, we
provide an intuitive picture that justifies \hspace*{2.1cm}some unusual features of the SR
model and proves its consistency.\vspace{0.5in}

\noindent
\hspace*{2.1cm}\textbf{KEY WORDS:} quantum mechanics, objectivity, realism, local\-\hspace*{2.1cm}ity, quantum
measurement, semantic realism.

\section{\normalsize{INTRODUCTION}}

One of us has recently proposed an \textit{SR model} which provides an
interpretation of quantum mechanics (QM) that is \textit{objective}.$^{(1)}$
Intuitively, objectivity means here that any
measurement of a physical property of an individual sample of a given
physical system reveals a preexisting value of the measured property, that
does not depend on the measurements that are carried out on the sample.%
\footnote{%
More rigorously, objectivity can be intended as a purely semantic notion, as
follows. Any physical theory is stated by means of a general language which
contains a \textit{theoretical language} L$_{T}$ and an \textit{observative
language} L$_{O}$. The former constitutes the formal apparatus of the theory
and contains terms denoting theoretical entities (as probability amplitudes,
electromagnetic fields, etc.). The latter is linked to L$_{T}$ by means of 
\textit{correspondence rules} that provide an indirect and partial
interpretation of L$_{T}$ on L$_{O}$. Furthermore, L$_{O}$ is interpreted by
means of \textit{assignment rules} which make some symbols of L$_{O}$
correspond to macroscopic entities (as preparing and measuring devices,
outcomes, etc.), so that the elementary sentences of L$_{O}$ are \textit{%
verifiable}, or \textit{testable}, since they state verifiable properties of
individual objects of the kind considered by the theory (note that this does
not imply that also the molecular sentences of L$_{O}$\ are testable). On
the basis of these assigments, a truth theory is (often implicitly) adopted
that defines truth values for some or all sentences of L$_{O}$. Then, we say
that physical properties are \textit{objective} in the given theory if the
truth values of \textit{all} elementary sentences of L$_{O}$ are defined
independently of the actual determination of them that may be done by an
observer (for instance, the correspondence theory of truth reaches this goal
by means of a set-theoretical model; by the way, we note that this truth
theory entails only a form of observative, or macroscopic, realism, even if
it is compatible with more demanding forms of realism).}

The SR model is inspired by a series of more general papers aiming to supply
an \textit{SR interpretation} of QM that is realistic in a semantic sense,
in the framework of an epistemological position called \textit{Semantic
Realism} (briefly, \textit{SR}; see, e.g., Refs. 2-5): indeed, it shows how
an SR interpretation can be consistently constructed. However, the SR model
is presented in Ref. 1 by using only the standard language of QM, in order
to make it understandable even to physicists that are not interested in the
conceptual subtelties of the general theory. But the treatment in Ref. 1
does not deal explicitly with the special case of compound physical systems,
hence neither the measurement problem nor the locality/nonlocality problem
are considered, even though the locality of the interpretation of QM
provided by the SR model is anticipated. Therefore, we intend to discuss
briefly these topics in the present paper.

Our analysis begins with some preliminaries. We discuss in Sec. 2 the
concept of physical property from a logical viewpoint, stress that
properties having different logical orders correspond to different kinds of
experimental procedures, and note that the properties represented by
projection operators in standard QM or in the SR model are first order
properties only; we also point out that, even if every state S of the
physical system can be associated with a first order property (the \textit{%
support} F$_{S}$ of S), recognizing an unknown state requires experimental
procedures corresponding to higher order properties, which is relevant to
the treatment of the measurement problem, as we show in Sec. 6. Furthermore,
we briefly analyze in Sec. 3 some typical proofs of nonlocality of standard
QM and individuate in them a common general scheme, notwithstanding their
differences.

Bearing in mind the above preliminaries, we deal with the locality problem
from the viewpoint of the SR model in Sec. 4. We provide firstly a slightly
simplified version of the model, and then note that the objective
interpretation of QM provided by it supplies an intuitive local picture of
the physical world and avoids a number of paradoxes, since objectivity
implies locality. But this entails that the arguments examined in Sec. 3
must fail to hold, otherwise one would get a contradiction. Thus, we
dedicate the rest of Sec. 4 to show that the proofs of nonlocality in Sec. 3
are actually invalid within the SR model, so that no inconsistency occurs.
As a byproduct of our analysis, we get that Bell's inequalities do not
provide a test for distinguishing local realistic theories from QM.

We then come to quantum measurements and observe in Sec. 5 that the SR model
avoids the main problem of standard quantum measurement theory, \textit{i.e}%
., the \textit{objectification} problem; we also note that measurements
still play a nonclassical role according to the SR model, since choosing a
specific measurement establishes which properties can be known and which
remain unknown, but point out some relevant differences between this
perspective and the standard QM viewpoint. Moreover, we show in Sec. 6 that
the further problem of double (unitary/stochastic) evolution of quantum
measurement theory disappears within the SR model, since stochastic
evolution can be considered as an approximate law that is valid \textit{for
all practical purposes}; we also discuss some consequences of the projection
postulate that illustrate further the differences existing between the
interpretation of the measuring process according to the SR model and the
standard interpretation.

Finally, we provide in Sec. 7 an intuitive picture that justifies some
relevant features of the SR model and proves its consistency by modifying
the \textit{extended SR model}$^{(1)}$ in which microscopic properties are
introduced as theoretical entities.

\section{\normalsize{PHYSICAL PROPERTIES, STATES AND SUPPORTS}}

Consider the following sets of statements in the standard language of
physics.

(i) ``The energy of the system falls in the interval [a,b]''.

``The system has energy $\mathcal{E}$\textit{\ }and momentum $\vec{p}$ at
time t''.

(ii) ``The energy of the system falls in the interval [a,b] with frequency 
\textit{f} whenever the system is in the state S''.

``If the system has energy $\mathcal{E}$, then its momentum is $\vec{p}$
with frequency \textit{f}''.

(iii) ``The energy of the system falls in the interval [a,b] with a
frequency that is maximal in the state S''.

``If the system has energy $\mathcal{E}$, then its momentum is $\vec{p}$
with a frequency that is maximal whenever the system is in the state S''.

All these statements express, in some sense, ``physical properties'' of a
physical system. But these properties have not the same logical status,
correspond to conceptually different experimental apparatuses, and a careful
analysis of their differences is useful if one wants to discuss the
objective interpretation of QM provided by the SR model in the case of
compound physical systems. Therefore, let us preliminarily observe that the
word \textit{system} in the above statements actually means \textit{%
individual sample of a given physical system}, or \textit{physical object}
according to the terminology introduced in the SR model (indeed the term 
\textit{physical system} is commonly used in the standard language of
physics for denoting both classes of physical objects and individual
samples, leaving to the context the charge of making clear the specific
meaning that is adopted). Then, let us note that the first statement in (i)
assigns the property

F=having energy that falls in the interval [a,b]

\noindent to a physical object, while the second statement assigns the
properties

E=having energy $\mathcal{E}$\textit{\ }at time \textit{t},

P=having momentum $\vec{p}$ at time \textit{t}.

The properties F, E, P are \textit{first order properties} from a logical
viewpoint, since they apply to individual samples, and each of them can be
tested (in a given laboratory) by means of a single measurement performed by
a suitable \textit{ideal} dichotomic registering device having outcomes 0
and 1 (of course, E and P can be tested conjointly only if they are
commeasurable).

Let us come to the statements in (ii). These assign \textit{second order
properties }to ensembles\textit{\ }of physical objects. To be precise, in
the first statement one considers the ensemble of objects that possess the
property F and the ensemble of objects that are in the state S, and the
second order property regards the number of objects in their intersection,
which must be such that its ratio with the number of the objects in S is 
\textit{f}. Analogously, in the second statement one considers the ensemble
of objects that possess the property E and the ensemble of objects that
possess the property P, and the second order property regards the number of
objects in their intersection, which must by such that its ratio with the
number of objects that possess the property E is \textit{f}. The first of
these properties can be tested by producing a given number of physical
objects in the state S, performing measurements of the first order property
F on its elements, counting the objects that have the property F, and then
calculating a relative frequency. The second property can be tested by means
of analogous procedures (which require measurements of first order
properties on a number of objects) if E and P are commeasurable, while there is
no procedure testing it in standard QM if E and P are not commeasurable.

Finally, the statements in (iii) assign \textit{third order properties} to 
\textit{sets of ensembles. }The property in the first statement can be
tested (in a given laboratory) by producing sets of ensembles, performing
measurements of first order properties on all elements of each ensemble,
calculating frequencies, and finally comparing the obtained results. The
property in the second statement requires analogous procedures, which may
exist or not, depending on the commeasurability of E and P.

It is now apparent that one could take into account further statements
containing properties of still higher order. Our discussion however is
sufficient to prove the main point here: properties of different logical
orders appear in the common language of physics, and properties that are
different when looked at from this logical viewpoint are also different from
a physical viewpoint. Of course, nothing prohibits that a first order
property F be attributed to some or all elements of an ensemble of physical
objects: but first order properties must be distinguished from higher order
properties, and, in particular, from \textit{correlation properties}, which
usually are second order properties that establish relations among first
order properties (the example above shows that the measurement of a property
of this kind requires the comparison of sets of results obtained by
measuring first order properties). We shall see that this distinction is
relevant when dealing with the measurement problem in Sec. 6.

From a mathematical viewpoint, only first order properties are represented
directly within standard QM. To be precise, let ($\mathcal{L}$($\mathcal{H}$%
),$\leq $) be the lattice of all orthogonal projection operators on the
Hilbert space $\mathcal{H}$ of a physical system, and let $\mathcal{L}$ be
the set of all first order properties of the system. According to standard
QM, every element of ($\mathcal{L}$($\mathcal{H}$),$\leq $) represents
bijectively (in absence of superselection rules) an element of $\mathcal{L}$%
. For the sake of brevity, we call any element of $\mathcal{L}$ \textit{%
physical property}, or simply \textit{property}, in the following, omitting
the reference to the logical order.

The set $\mathcal{L}$ can be endowed with the partial order induced on it by
the mathematical order $\leq $ defined on $\mathcal{L}$($\mathcal{H}$) (that we still
denote by $\leq $), and the lattice ($\mathcal{L}$,$\leq $) is usually
called \textit{the lattice of properties} of the system. It follows that
every pure state S can be associated with a \textit{minimal} property F$%
_{S}\in \mathcal{L}$ that is often called \textit{the support} of S in the
literature (see, e.g., Ref. 6). To be precise, if S is represented in $%
\mathcal{H}$ by the vector $\mid \varphi \rangle $, F$_{S}$ is the property
represented by the one-dimensional projection operator P$_{\varphi }$ = $%
\mid \varphi \rangle \langle \varphi \mid $, which obviously is such that P$%
_{\varphi }$ $\leq $ P for every P $\in \mathcal{L}$($\mathcal{H}$) such
that $\mid \mid $P$\mid \varphi \rangle \mid \mid ^{2}$= 1. It is then
apparent that F$_{S}$ can be characterized as the property that is possessed
by a physical object x \textit{with certainty} (\textit{i.e}., with
probability 1) iff x is in the state S. Indeed, for every vector $\mid
\varphi ^{\prime }\rangle $ representing a pure state S$^{\prime }$, one
gets $\mid \mid $P$_{\varphi }\mid \varphi ^{\prime }\rangle \mid \mid ^{2}$%
= 1 iff $\mid \varphi ^{\prime }\rangle $ = $e^{i \theta} \mid \varphi \rangle $, hence
iff S$^{\prime }$ = S.

The existence of a support for every pure state of a physical system is
linked with the problem of distinguishing different pure states, or pure
states from mixtures, in standard QM. Indeed, there is no way in this theory
for recognizing experimentally the state S of a single physical object x
whenever this state is not known (for the sake of brevity, we assume here
that S is a pure state): even if one measures on x an observable $\mathcal{A}
$ that has S as an eigenstate corresponding to a nondegenerate eigenvalue 
\textit{a}, and gets just \textit{a} (equivalently, if one tests the support
F$_{S}$ of S and gets that F$_{S}$ is possessed by x), one cannot assert
that the state of x was S before the measurement, since there are many
states that could yield outcome \textit{a} and yet are different from S (for
instance, all pure states that are represented by vectors that are not
orthogonal to the vector representing S). But if one accepts the definition
of states as equivalence classes of preparing devices propounded by Ludwig
(and incorporated within the SR model$^{(1)}$) one can know whether a given
preparing device $\pi $ prepares physical objects in the state S (briefly,
one can recognize S) by measuring mean values of suitable observables, which
is obviously equivalent to testing second order properties. The simplest way
of doing that is testing F$_{S}$ on a huge ensemble of objects prepared by $%
\pi $ by means of an ideal dichotomic device \textit{r}: indeed, one can
reasonably assume that $\pi $ belongs to the state S whenever \textit{r}
yields outcome 1 on \textit{all} samples, that is, whenever F$_{S}$ is
possessed by every physical object x prepared by $\pi $ or, equivalently,
the mean value of F$_{S}$ is 1. In particular, if S is an entangled state of
a compound physical system made up by two subsystems, this procedure allows
one to distinguish S from a mixture M$_{S}$\ corresponding to S via
biorthogonal decomposition (see, e.g., Ref. 7). Also this remark is relevant
to the quantum theory of measurement (see Sec. 6).

$\,$

\section{\normalsize{NONLOCALITY WITHIN STANDARD QM}}

The issue of nonlocality of QM was started by a famous paper by Einstein,
Podolski and Rosen (EPR),$^{(8)}$ which however had different goals: indeed,
it aimed to show that some reasonable assumptions, among which locality,
imply that standard QM is not complete (in a very specific sense introduced
by the authors), hence it can not be considered as a final theory of
microworld. Later on, the thought experiment proposed by EPR, regarding two
physical systems that have interacted in the past, was reformulated by Bohm$%
^{(9)}$ and a number of further thought experiments inspired by it were
suggested and used in order to point out the conflict between standard QM
and locality. Hence, one briefly says that standard QM is a nonlocal theory.

As anticipated in Sec. 1, we want to schematize some typical proofs of
nonlocality in this section, in order to prepare the ground to our criticism
in Sec. 4. For the sake of clearness, we proceed by steps.

\smallskip

(1) The existing proofs of nonlocality of QM can be grouped in two classes
(see, e.g., Ref 10). $^{(i)}$The proofs showing that \textit{deterministic
local theories} are inconsistent with QM. $^{(ii)}$The more general proofs
showing that \textit{stochastic local theories} (which include deterministic
local theories) are inconsistent with QM. For the sake of brevity, we will
only consider the proofs in $^{(i)}$. It is indeed rather easy to extend our
analysis and criticism to the proofs in $^{(ii)}$.

\smallskip

(2) We denote by QPL in the following a set of \textit{empirical} quantum
laws, which may be void (intuitively, a physical law is empirical if it can
be directly checked, at least in principle, by means of suitable
experiments, such as, for instance, the relations among compatible
observables mentioned in the KS condition;$^{(1)}$ a more precise
distinction between empirical and \textit{theoretical} laws will be
introduced in Sec. 4, (2)). We denote by LOC the assumption that QM is a
local theory (in the standard EPR sense,\footnote{%
``Since at the time of measurement the two systems no longer interact, no
real change can take place in the second system in consequence of anything
that can be done to the first system''.$^{(8)}$} that can be rephrased by
saying that a measurement on one of many spatially separated subsystems of a
compound physical system does not affect the properties of the other
subsystems). Finally, we denote by R the following assumption.

R. \textit{The values of all physical properties of any physical object are
predetermined for any measurement context.}

\smallskip

(3) Bearing in mind the definitions in (2), the general scheme of a typical
proof of nonlocality is the following.

Firstly, one proves that

QPL \textit{and} LOC \textit{and} R $\Rightarrow $ (\textit{not} QM),

\noindent or, equivalently,

QM $\Rightarrow $ (\textit{not} QPL) \textit{or} (\textit{not} LOC) \textit{%
or} (\textit{not} R).

Secondly, since QM $\Rightarrow $ QPL, one gets

QM $\Rightarrow $ (\textit{not} LOC) \textit{or} (\textit{not} R).

Finally, one proves that

QM \textit{and} (\textit{not} R) $\Rightarrow $ (\textit{not} LOC),

\noindent so that one concludes

QM $\Rightarrow $ (\textit{not} LOC).

\smallskip

(4) Let us consider some proofs of nonlocality and show that they actually
follow the scheme in (3).

\textit{Bell's original proof}.$^{(11)}$ Here, the Bohm variant of the EPR
thought experiment is considered (which refers to a compound system made up
by a pair of spin-1/2 particles formed somehow in the singlet spin state and
moving freely in opposite directions). Then, a \textit{Bell's inequality}
concerning some expectation values (hence a physical law linking second
order properties, see Sec. 2) is deduced by using assumptions LOC and R
together with a \textit{perfect correlation law} (PC: if the measurement on
one of the particles gives the result \textit{spin up} along the \textit{u}
direction, then a measurement on the other particle gives the result \textit{%
spin down} along the same direction), which is an empirical law linking
first order properties and following from the general theoretical laws of
QM. The deduction is based on the fact that assumption R allows one to
introduce \textit{hidden variables} specifying the state of a physical
system in a more complete way with respect to the quantum mechanical state.
Then, the expectation values predicted by QM are substituted in the Bell's
inequality and found to violate it.

The above procedure can be summarized by the implication PC \textit{and} LOC 
\textit{and} R $\Rightarrow $ (\textit{not} QM), which matches the first
step in the general scheme, with PC representing QPL in this particular
case. One thus obtains QM $\Rightarrow $ (\textit{not} LOC) \textit{or} (%
\textit{not} R) and concludes that QM contradicts local realism. The last
step in the scheme was not done explicitly by Bell and can be carried out by
adopting, for instance, the proof that PC \textit{and} LOC $\Rightarrow $ R
(hence PC\ \textit{and} (\textit{not} R) $\Rightarrow $ (\textit{not} LOC))
propounded by Redhead.$^{(12)}$

We add that the same paradigm, with PC as a special case of QPL, occurs in
different proofs, as Wigner's$^{(13)}$ and Sakurai's.$^{(14)}$ In these
proofs, however, an inequality is deduced (still briefly called \textit{%
Bell's inequality}) that concerns probabilities rather than expectation
values.

\textit{Clauser et al.'s proof}.$^{(15)}$ This proof introduces a
generalized Bell's inequality, sometimes called \textit{BCHSH's inequality},
that concerns expectation values (hence it expresses a physical law linking
second order properties, as Bell's inequality). This inequality is compared
with the predictions of QM, finding contradictions. BCHSH's inequality is
deduced by using LOC and R only, so that one proves that LOC \textit{and} R $%
\Rightarrow $ (\textit{not} QM), hence QPL is void in this case. The rest of
the proof can be carried out as in the general scheme.

\textit{Greenberger et al.'s proof}.$^{(16)}$ Here no inequality is
introduced. A system of four correlated spin-1/2 particles is considered,
and the authors use directly a perfect correlation law PC$_{1}$ (that
generalizes the PC law mentioned above), R and LOC\footnote{%
To be precise, the authors introduce, besides LOC, \textit{realism} and 
\textit{completeness} in the EPR sense. These assumptions are however
equivalent, as far as the proof is concerned, to assumption R.} in order to
obtain a contradiction with another perfect correlation law PC$_{2}$, hence
with QM. Thus, the authors prove that PC$_{1}$ \textit{and} LOC \textit{and}
R $\Rightarrow $ (\textit{not} QM), which matches the first step in the
general scheme, with PC$_{1}$ representing QPL in this case. Again, the rest
of the proof can be carried out as in the general scheme.

\textit{Mermin's proof}.$^{(17)}$ Also this proof does not introduce
inequalities. The author takes into account a system of three different
spin-1/2 particles, assumes a quantum physical law linking first order
properties (the product of four suitably chosen dichotomic nonlocal
observables is equal to -1) together with LOC and (implicitly) R, and shows
that this law cannot be fulfilled together with other similar laws following
from QM. Thus, also Mermin proves an implication of the form QPL \textit{and}
LOC \textit{and} R $\Rightarrow $ (\textit{not} QM), from which the argument
against LOC can be carried out as in the general scheme.

\smallskip

(5) The analysis in (4) shows that the scheme in (3) provides the general
structure of the existing proofs of nonlocality. In this scheme, assumptions
R and LOC play a crucial role. Let us therefore close our discussion by
comparing R and LOC with the assumption of objectivity (briefly, O), which
plays instead a crucial role in the proofs of contextuality of standard QM.$%
^{(5)}$

Let us note firstly that assumption R expresses a minimal form of realism.
This realism can be meant in a purely semantic sense, as objectivity (see
Sec. 1), hence it is compatible with various forms of ontological realism
(as the assumption about the existence of elements of reality in the EPR
argument\footnote{%
``If, without in any way disturbing a system, we can predict with certainty (%
\textit{i.e}., with probability equal to 1) the value of a physical
quantity, then there exists an element of physical reality corresponding to
this physical quantity''.$^{(8)}$}) but does not imply them.$^{(4)}$ Yet, R
is weaker than O. Indeed, O entails that the values of physical properties
are independent of the measuring apparatuses (\textit{noncontextuality}),
while R may hold also in a contextual theory (as Bohm's), since it requires
only that the values of physical properties are not brought into being by
the very act of measuring them.\footnote{%
In order to avoid misunderstandings, we note explicitly that assumption R
coincides with assumption R in Ref. 4 and not with assumption R in Ref. 5,
which is instead the assumption of objectivity.}

Let us note then that O also implies LOC, since it entails in particular
that the properties of the subsystems of a compound physical system exist
independently of any measurement. By putting this implication together with
the implication O $\Rightarrow $ R, one gets O $\Rightarrow $ LOC \textit{and%
} R. However, the converse implication does not hold, since R and LOC are
compatible with the existence of measurements that do not influence each
other at a distance but influence locally the values of the properties that
are measured. Thus, we conclude that R and LOC are globally weaker than O.

$\,$

\section{\normalsize{RECOVERING LOCALITY WITHIN THE SR MODEL}}

It is well known that nonlocality of standard QM raises a number of problems
and paradoxes. However, it has been proven in several papers (see, e.g.,
Refs. 4 and 18-20) that the general SR interpretation of QM invalidates some
typical proofs of nonlocality. Basing on our analysis in Sec. 3, we want to
attain in this section a similar result within the framework of the SR
model, which has the substantial advantage of avoiding a number of logical
and epistemological notions, making things clear within the standard
language of QM. To this end, we use throughout in the following the
definitions and concepts introduced in Ref. 1.

It is important to observe that we are led to question the proofs of
nonlocality not only for general theoretical reasons (in particular,
attaining a more intuitive \textit{local} picture of the physical world) but
also to avoid inconsistencies. Indeed, as we have anticipated in Sec. 1, the
proofs in Sec. 3 raise a consistency problem. For, should they be valid
within the SR model, QM would be a nonlocal theory also according to this
model: but, then, the model could not provide an objective interpretation of
QM, since objectivity entails both R and LOC (see (5) in Sec. 3).

For the sake of clearness, we again proceed by steps.

\smallskip

(1) Let us summarize some features of the SR model and introduce in it some
simplifications that make it more intuitive and easy to handle.

Firstly, the main feature of the model is the substitution of every physical
observable $\mathcal{A}$ of standard QM with an observable $\mathcal{A}_{0}$
in which a \textit{no-registration outcome} a$_{0}$ is added to the spectrum
of $\mathcal{A}$. This is an old idea. Yet, a$_{0}$ is interpreted in the
model as a possible result of a measurement of $\mathcal{A}_{0}$ providing
information about the physical object that is measured, which introduces a
new perspective, since the occurrence of a$_{0}$ is usually attributed to a
lack of efficiency of the measuring apparatus. The family of properties
associated with an observable $\mathcal{A}_{0}$ is then $\mathcal{E}_{%
\mathcal{A}_{0}}$ = $\left\{ \text{(}\mathcal{A}_{0}\text{,}\Delta \text{)}%
\right\} _{\Delta \in \mathcal{B}(\mathbb{R})\text{)}}$ (where $\mathcal{B}$(%
$\mathbb{R}$) denotes the $\sigma $-ring of all Borel sets on the real line $%
\mathbb{R}$).

Secondly, states are defined as in standard QM, and all properties are
assumed to be objective, hence the information about a physical object x
following from the knowledge of its state is incomplete. Thus, there is a
set of properties that are possessed by all objects in the state S (the set
of properties that are\textit{\ certainly true} in S, see Sec. 5), but
different objects in S may differ because of further properties. One can
then focus on some additional properties, take them (together with S) as
initial, or boundary, conditions, and consider the subset of all objects in
S that possess them. Different choices of the additional properties lead to
different subsets (which may intersect). We briefly say that one can
consider different \textit{physical situations}. It is then apparent that
these situations can be partitioned in two basically different classes.
Indeed, one can choose as additional properties the property of being
detected \textit{and} some further properties (possibly none) that are
pairwise commeasurable (\textit{i.e.}, simultaneously measurable, see Ref.
1): in this case an \textit{accessible }physical situation is considered. On
the contrary, if one chooses the property of being not detected \textit{or}
some further properties that are not pairwise commeasurable, a \textit{%
nonaccessible }physical situation is considered.

Thirdly, the following nonclassical \textit{Metatheoretical Generalized
Principle} holds within the SR model (the same principle was stated in a
number of previous papers, basing on general arguments, see, e.g., Refs. 3
and 5).

\noindent MGP. \textit{A physical statement expressing an empirical physical
law is true whenever an accessible physical situation is considered, but it
may be false (as well as true) whenever a nonaccessible physical situation
is considered}.

Finally, properties are represented by projection operators, as in standard
QM, but the representation is not one-to-one, since the property ($\mathcal{A%
}_{0}$,$\Delta $), with a$_{0}\notin \Delta $, and the property ($\mathcal{A}%
_{0}$,$\Delta \cup \left\{ \text{a}_{0}\right\} $) are represented by the
same operator. This feature, however, makes the SR model unnecessarily
complicate, since the representation of any property ($\mathcal{A}_{0}$,$%
\Delta $), with a$_{0}\in \Delta $, is not needed in the following (nor to
reach the conclusions in Ref. 1). Thus, we modify it here by simply assuming
that projection operators represent bijectively (up to a physical
equivalence relation, see Ref. 1) all properties of the form ($\mathcal{A}%
_{0}$,$\Delta $), with a$_{0}\notin \Delta $, while all remaining
properties, though entering physical reasonings, have no mathematical
representation. Then, we assume that the probability of a given property of
the form ($\mathcal{A}_{0}$,$\Delta $), with a$_{0}\notin \Delta $, for a
physical object in a given state can be evaluated, \textit{in every
accessible physical situation}, by referring to the representation of states
and properties and using the rules of standard QM (the name \textit{SR model 
}will refer to this simplified version of the model from now on). It follows
in particular that all quantum laws expressed by the standard mathematical
language of QM link only properties such that a$_{0}\notin \Delta $, which
becomes intuitively clear within the picture provided at the end (see Sec.
7).

\smallskip

(2) MGP is an \textit{epistemological} principle regarding the range of
validity of physical laws, and it implies a change in our way of looking at
the laws of QM, not a change in the laws themselves. Moreover, it provides
the main argument against the standard proofs of nonlocality, that are thus
criticized from an epistemological rather than from a technical viewpoint.
MGP plays therefore a crucial role within the SR model, and it may be useful
to reconsider here some arguments that suggest introducing it.

Let us begin by making clear the distinction between \textit{empirical} and 
\textit{theoretical} physical laws that is presupposed by MGP. In any
theory, hence in QM, a law is said to be empirical if it is obtained from
theoretical laws of the formal apparatus of the theory via correspondence
rules and is expressed by a \textit{testable} sentence of the observative
language of the theory (see footnote 1), so that it may undergo a process of
empirical control (theoretical laws can then be seen as \textit{schemes of
laws}, from which empirical laws can be deduced$^{(21)}$). This definition,
however, does not mean that an empirical law can always be checked. Let us
consider, for example, an empirical law in which only pairwise commeasurable
first order properties appear. If x is a physical object (in a state S) that
is detected and, furthermore, additional properties are chosen that
are pairwise commeasurable, and these properties are pairwise commeasurable
with all properties appearing in the law, then an accessible physical
situation is considered (see (1)), and one can check both whether x actually
has all the hypothesized properties and whether the law holds. On the
contrary, whenever x is not detected, or it is detected but additional
properties are chosen that are not pairwise commeasurable with the
properties appearing in the law, a nonaccessible physical situation is
considered, and it is impossible either to make a check (if x is not
detected) or to check both whether x has the hypothesized properties and
whether the law holds (if x is detected).

The above example shows that one can consider physical situations in which
an empirical physical law cannot, in principle, be checked. It is then
consistent with the operational philosophy of QM to assume that the validity
of an empirical law can be asserted only in accessible physical situations,
which directly leads to MGP.

Finally, we note that, whenever an empirical law linking physical properties
is expressed by means of the standard mathematical language of QM, then only
objects that are actually detected are automatically considered, because of
the simplification of the model introduced in (1). Also in this case,
however, nonaccessible physical situations can occur whenever properties
appear in the initial conditions (see (1)) that are represented by
projection operators which do not commute with the projection operators
representing the properties that appear in the law.

\smallskip

(3) Let us come now to our criticism of the standard proofs of nonlocality.
Because of our analysis in Sec. 3, (4), this criticism can be carried out by
referring to the scheme in Sec. 3, (3). Let us firstly show that the
standard proofs of nonlocality are doubtful even if the SR model is ignored.
To this end, note that also assumption R, though weaker than objectivity,
implies that nonaccessible physical situations can be considered. Indeed, it
follows from R that a value is defined for every property of a physical
object x and for every measurement context, so that it is sufficient to choose
a set of noncommeasurable properties for considering a situation of this
kind. Moreover, direct inspection shows that the scheme in Sec. 3, (3) is
not exhaustive. Indeed, all proofs of nonlocality considered in Sec. 3, (4)
use, besides the explicit assumptions stated in the scheme, the implicit
assumption that \textit{all empirical quantum laws can be applied to
physical objects in every physical situation, be it accessible or not}. This
assumption is problematic if viewed at from a standard operational quantum
viewpoint, since it introduces within QM a classical conception of empirical
physical laws (we therefore call it \textit{Metatheoretical Classical
Principle} or briefly MCP,$^{(3)}$ in the following\footnote{%
Note that MCP implies assuming the \textit{KS condition} stated by Kochen
and Specker ``for the successful introduction of hidden variables'' in their
proof of contextuality of QM,$^{(22)}$ and then adopted in all successive
proofs of contextuality. Hence, our present criticism generalizes the
criticism to the KS condition carried out in Ref. 1.}).

Let us come, however, to the SR model. Within this model, objectivity
implies that accessible and nonaccessible physical situations must be
introduced, and MCP can be proven to break down by means of an example,
while the weaker MGP holds.$^{(1)}$ Thus, the proofs in Sec. 3, (4) are
invalid and the consistency problem pointed out at the beginning of this
section is avoided.

To complete our argument, it remains to point out where exactly each of the
aforesaid proofs uses MCP. Let us discuss this issue in some details.

\smallskip

(4) Firstly, let us consider Bell's,$^{(11)}$ Wigner's$^{(13)}$ and Sakurai's%
$^{(14)}$ proofs. Here, it is immediate to see that the PC law is applied
repeatedly to physical objects that are hypothesized to possess spin up along
non-parallel directions. Hence, this law is directly applied in
nonaccessible physical situations, implicitly adopting MCP and violating MGP.

Secondly, let us consider Clauser's \textit{et al}.'s$^{(15)}$ proof. This
proof deserves special attention, since no physical law linking first order
properties is used in it. Therefore, let us observe that the BCHSH inequality contains a sum of expectation values in which
noncompatible observables occur. Hence, every expectation value must be
evaluated making reference to different sets of physical objects: all
objects are prepared in the same entangled state (to be precise, the singlet
state), but in each set only commeasurable physical properties are measured,
which differ from set to set. The same procedure is needed if the quantum
inequality corresponding to the BCHSH inequality is considered. Yet,
according to the SR model, the expectation values in the BCHSH inequality
are evaluated taking into account \textit{all} physical objects in each set.
On the contrary, the quantum rules provide probabilities referring to
accessible situations only,$^{(1)}$ hence the expectation values in the
quantum inequality take into account only the subsets of objects that are
actually detected in each set. These subsets are selected by apparatuses
that differ from set to set, and could be unfair statistical samples of the
whole set to which they belong, since in each set the measurements select,
because of the no-registration outcome that can occur,$^{(1)}$ a subset of
physical objects in which the statistical relations among the measured
properties can be different from the relations that hold in the original set
and from the relations that hold in the other sets. Thus, the BCHSH and the
quantum inequality cannot be identified, and no contradiction with QM occurs.

Note that we have avoided using MGP in the above argument since MGP has not
been justified explicitly in the case of empirical laws linking second order
properties (see (2); it is interesting to observe that the argument can then
be used in order to justify MGP in this case). But if one accepts MGP, one
can simply say that the quantum predictions on probabilities can be invalid,
because of MGP, if also physical objects in nonaccessible physical
situations are considered, as in the BCHSH inequality: hence, identifying
this inequality with the corresponding quantum inequality violates MGP.

Thirdly, let us consider Greenberger \textit{et al}.'s$^{(16)}$ proof. Here,
different perfect correlation laws PC$_{1}$ and PC$_{2}$ are applied to the
same physical object, and the properties in PC$_{1}$ are not all
commeasurable with the properties in PC$_{2}$. Hence a nonaccessible
physical situation is envisaged in which PC$_{1}$ and PC$_{2}$ are assumed
to be valid, implicitly adopting MCP and violating MGP.

Finally, let us consider Mermin's$^{(17)}$ proof. Here, different
quantum laws are applied to a given physical object, and there are
observables in some of the laws that are not compatible (in the standard
sense of QM) with some observables in the other laws. From the viewpoint of
the SR model, this produces a nonaccessible physical situation in which some
empirical physical laws are assumed to be valid, implicitly adopting MCP and
violating MGP.

\smallskip

(5) Coming back to the scheme in Sec. 3, (3), our arguments in (3) and (4) above 
can be summarized by saying that the implication QPL \textit{and} LOC 
\textit{and} R $\Rightarrow $ (\textit{not} QM) must be substituted by QPL 
\textit{and} LOC \textit{and} R \textit{and} MCP $\Rightarrow $ (\textit{not}
QM), which does not hold within the SR model (and is criticizable even
within standard QM, see (3)), so that (\textit{not} LOC) does not follow
from QM.

We note explicitly that one can still object that our arguments in (3) are
not conclusive. Indeed, objectivity does not imply only R and LOC, as
pointed out in Sec. 3, (5), but also the weaker assumptions that are
introduced when considering local stochastic theories (as the factorization
of probabilities in \textit{objective local theories}, see, e.g., Ref. 10),
so that one should still show that the proofs that these theories are
inconsistent with QM are invalid within the SR model. As we have anticipated
in Sec. 3, (1), however, it is easy to extend our arguments in (3) in order
to attain this invalidation.

\smallskip

(6) The result obtained in (3) raises, in particular, the problem of the
role of Bell's inequalities. Indeed, these inequalities are usually
maintained to provide crucial tests for discriminating between local realism
and QM (see, e.g., Refs. 7 and 23). One may then wonder about what would
happen, according to the SR model, if one should perform a test of a Bell's
inequality: would it be violated or not? The answer is that there are
basically two kinds of Bell's inequalities, as our analysis in Sec. 3, (4)
shows. Those obtained from assumptions LOC and R only, as BCHSH's
inequality, are correct theoretical formulas which are not epistemically
accessible, hence cannot be tested (see also Refs. 4 and 24, where however
the original Bell's inequality was not classified properly). Any physical
experiment tests something else (correlations among commeasurable properties
of physical objects that are actually detected), hence yields the results
predicted by QM. No contradiction occurs, since the inequalities that can be
tested in QM could be identified with Bell's inequalities only violating
MGP. Thus, the latter inequalities do not provide methods for testing
experimentally whether either QM or local realism is correct, according to
the SR model. But the difference between quantum inequalities and Bell's
inequalities proves that some quantum laws regarding compound systems fail
to hold in nonaccessible physical situations.

The above arguments do not apply to the inequalities that are deduced by
using repeatedly a non void set QPL of empirical quantum laws besides R and
LOC. These inequalities are simply incorrect according to the SR model,
since they are deduced by applying QPL in nonaccessible physical situations.

$\,$

\section{\normalsize{OBJECTIVITY AND MEASUREMENT}}

The objective interpretation of QM provided by the SR model avoids from the
very beginning the main problem of the quantum theory of measurement, 
\textit{i.e}. the \textit{objectification problem}. Indeed, it allows one to
interpret a measurement of a property F on a physical object x as an inquiry
about whether F is possessed or not by x, not as an objectification of F.
Generally speaking, this brings back the measurement problem to classical
terms.

It must be stressed, however, that some typical quantum features do not
disappear in the SR model. In particular, the existence of a non-trivial
commeasurability relation prohibits one from testing all properties
possessed by a given physical object x conjointly, so that the knowledge of
all properties of x in a given state cannot be provided by any measurement.
It is thus interesting to inquire further into the knowledge of properties
that one attains when the state of x is specified. Therefore, let us suppose
that x is in the (pure) state S represented by the vector $\mid \varphi
\rangle $.\footnote{%
Because of the projection postulate, in standard QM one can obtain physical
objects in the state S by choosing an observable $\mathcal{A}$ that has S as
eigenstate belonging to a nondegenerate eigenvalue $a$, performing ideal measurements
of $\mathcal{A}$ and selecting the objects that yield the outcome $a$. This
procedure (with $\mathcal{A}_{0}$ in place of $\mathcal{A}$) is still valid
within the SR model (see Sec. 6, Eq. (1)).} It is easy to see that in the
SR model (as in standard QM) a subset $\mathcal{E}_{S}$ exists, made up by
properties that are \textit{certainly true} in S (that is, have probability
1 in S). To be precise, $\mathcal{E}_{S}$ contains all properties of the
form ($\mathcal{A}_{0}$,$\Delta $) such that:

(i) $\mathcal{A}_{0}$ is a physical observable obtained from an observable $%
\mathcal{A}$ of standard QM by adding a no-registration outcome a$_{0}$, see
Sec. 4, (1);

(ii) $\Delta $ is a Borel set on the real line $\mathbb{R}$ that includes a$%
_{0} $;

(iii) ($\mathcal{A}_{0}$,$\Delta \backslash \left\{ \text{a}_{0}\right\} $)
is represented by a projection operator P on the Hilbert space $\mathcal{H}$
of the system such that P$\mid \varphi \rangle $ = $\mid \varphi \rangle $.

\noindent Indeed, if $\mathcal{A}_{0}$ is measured, one either gets outcome a%
$_{0}$ (x is not registered) or an outcome that belongs to $\Delta
\backslash \left\{ \text{a}_{0}\right\} $, since standard quantum rules hold
for evaluating the probability of ($\mathcal{A}_{0}$,$\Delta \backslash
\left\{ \text{a}_{0}\right\} $) in the state S (see Sec. 4, (1)), and $\Vert 
$P$\mid \varphi \rangle \Vert ^{2}$ = 1 because of (iii). Analogously, one
sees that a subset $\mathcal{E}_{S}^{\bot }$ made up by properties that are 
\textit{certainly false} in S (that is, have probability 0 in S) exists. To
be precise, $\mathcal{E}_{S}^{\bot }$ contains all properties of the form ($%
\mathcal{A}_{0}$,$\Delta $) such that $\mathcal{A}_{0}$ is defined as in
(i), $\Delta $ is a Borel set on the real line that does not include a$_{0}$%
, and ($\mathcal{A}_{0}$,$\Delta $) is represented by a projection operator
P such that P$\mid \varphi \rangle $ = 0. Thus, one obtains that the
specification of the state S of x provides information about the properties
in the set $\mathcal{E}_{S}\cup \mathcal{E}_{S}^{\bot }$.

The set $\mathcal{E}_{S}\cup \mathcal{E}_{S}^{\bot }$, however, is strictly
contained in the set $\mathcal{E}$ of all properties, each of which is
objective according to the SR model. It is indeed easy to see that there are
properties in $\mathcal{E}$\TEXTsymbol{\backslash}($\mathcal{E}_{S}\cup 
\mathcal{E}_{S}^{\bot }$) such that one cannot deduce from knowing S whether
they are possessed or not by x. Hence, the information provided by S is
incomplete within the SR model (see Sec. 4, (1)). This has many relevant
consequences. Let us point out here some of them.

Firstly, some pairwise commeasurable properties in $\mathcal{E}$\TEXTsymbol{%
\backslash}($\mathcal{E}_{S}\cup \mathcal{E}_{S}^{\bot }$) can be tested, so
that for each of them one can say whether it was possessed or not by x (in
the state S) before the measurement, increasing our information on x without
introducing contradictions. This possibility does not occur in standard QM,
where a property that has not probability 1 or 0 is not real in S, and a
test actualizes it, so that one can say that it is possessed or not by x
only after the measurement (hence, in general, in a state that is different
from S). Thus, conjoint knowledge of pairs of arbitrary properties can be
obtained in some cases in the SR model. For example, whenever one can
predict that x possesses a property F$_{1}$ at time t and a measurement of
another property F$_{2}$ on x at t yields outcome 1, then one knows that
both F$_{1}$ and F$_{2}$ are possessed by x at t, whatever F$_{1}$ and F$%
_{2} $ may be.

Secondly, note that the conjoint knowledge mentioned above does not survive,
in general, after t, since the interaction between x and the measuring
apparatus may change the state of x (this issue will be discussed in more
details in Sec. 6, referring to the special case of ideal measurements).
Whatever it may be, however, such a change does not imply that the
properties of x must be different after the measurement: it only implies a
modification of our knowledge about the properties that are certainly
possessed or not possessed by x. In different words, \textit{a measurement
changes our information about x, but does not change necessarily the
properties of x} (though a change of properties may occur because of the
interaction with the measuring apparatus). Again, this perspective is
different from the perspective of standard quantum measurement theory, in
which a change of state implies a change of the properties that are actual
for x. We briefly call \textit{epistemic conception of states} in the
following the new viewpoint introduced by the SR model.

$\,$

\section{\normalsize{THE PROJECTION POSTULATE}}

Our discussion of measurements within the SR model in Sec. 5 is carried out
by considering the microscopic physical system and the macroscopic measuring
apparatus as distinct physical entities, according to a standard elementary
way of dealing with measurement processes. It is well known, however, that
some crucial problems occur in standard QM whenever one tries to select the
subclass of apparatuses performing \textit{ideal} (repeatable) measurements
and treat them as macroscopic quantum systems, in order to provide a more
complete description of the measurement process by considering the compound
system formed by the microscopic system plus the measuring apparatus within
standard QM. Indeed, two major difficulties occur.

(i) The unitary evolution of the whole system predicted by the Schr\"{o}%
dinger equation implies that a pure initial state evolves into a pure final
state, which may be entangled in such a way that neither the microscopic
system nor the macroscopic apparatus possess individual properties.\footnote{%
We remind that, in order to know whether this occurs, one can consider the
biorthogonal decomposition of the vector $\mid \chi \rangle $ representing
the final state S, according to which $\mid \chi \rangle $ = $\sum_{i\in I}%
\sqrt{p_{i}}$ $\mid \varphi _{i}\rangle \mid \psi _{i}\rangle $, with 
\textit{I} a set of indexes, \textit{p}$_{i}>0$, $\sum_{i\in I}p_{i}=1$, $%
\mid \varphi _{i}\rangle $ and $\mid \psi _{i}\rangle $ vectors representing
states of the microscopic object and macroscopic apparatus, respectively.
Whenever $\left\{ \mid \varphi _{i}\rangle \right\} _{i\in I}$ and $\left\{
\mid \psi _{i}\rangle \right\} _{i\in I}$ are bases in the Hilbert spaces of
the two component subsystems, it is easy to verify (by considering any
projection operator representing a physical property of one of the two
subsystems) that neither of these may possess individual properties. By the
way, we also remind that the biorthogonal decomposition allows one to
associate a state M$_{S}$, which is represented by the density operator $%
\rho =\sum_{i\in I}p_{i}$ $\mid \varphi _{i}\rangle \mid \psi _{i}\rangle
\langle $ $\varphi _{i}\mid \langle \psi _{i}\mid $, with the pure state S.
Of course, M$_{S}$ is a mixture (and differs from S) if \textit{I} contains
more than an element.}

(ii) The experimental situation at the end of a measurement is described in
standard QM by the projection postulate, which can be justified whenever the
component subsystems are considered separately.\footnote{%
By using the symbols in footnote 8, we remind that this justification can be
attained by representing the final state S by means of the projection
operator P$_{S}=\mid \chi \rangle \langle \chi \mid $ rather than by means
of the vector $\mid \chi \rangle $, and then performing a partial trace of P$%
_{S}$ with respect to the subsystem that one does not want to consider.}
Yet, the projection postulate leads to predict a \textit{stochastic evolution%
} according to which the final state of the whole system is a mixture rather
than a pure state (to be precise, the mixture corresponding to the final
state in (i) via biorthogonal decomposition$^{8}$). This prediction implies
that the component subsystems possess individual properties, which is
consistent with observative data (one can indeed observe properties of the
measuring apparatus directly). But it is then unclear how the stochastic
evolution can be reconciled with the unitary evolution that should occur
according to the Schr\"{o}dinger equation, and, in particular, with
nonobjectivity.

The attempts to solve the above problems have produced a huge and generally
known literature, that we do not list here for the sake of brevity. Rather,
we would like to provide in this section a first, qualitative treatment of
these problems from the viewpoint of the SR model.

First of all, we note that we have not yet introduced any assumption about
time evolution in the SR model. However, consistency with standard QM
suggests one to maintain that also in this model the vector representing a
pure state of the whole system undergoes unitary evolution. Furthermore, one
is also led to maintain that, whenever ideal measurements are performed and
the a$_{0}$ outcome does not occur, the projection postulate provides a good
description of probabilities and final states of the system.

Let us come now to difficulties (i) and (ii) above. It is apparent that the
assumption of unitary evolution, though identical to the standard QM
assumption, does not raise any problem within the SR model when applied to
the measurement process. Indeed, objectivity implies that every conceivable
property of microscopic system or measuring apparatus either is possessed or
not by the subsystem that is considered, even if one cannot generally know 
\textit{a priori} which of the two alternatives occurs (it follows, in
particular, that no special argument, or additional assumption, or
modification of time evolution is needed in order to explain macroscopic
objectivity). This implies that difficulty (i) disappears. Furthermore, also
the contradiction in (ii) between unitary and stochastic evolution is far
less relevant because of objectivity, since all properties of the component
subsystems are actual according to both descriptions within the SR model.
Hence, one can safely resort to the old idea of reconciling the two
descriptions by assuming that one of them is theoretically rigorous, the
other one is approximate. In order to implement this idea, let us provide a
possible scheme of description of the measuring process within the SR model,
matching the standard simplified description of this process that is used in
many books in order to show in an elementary way that unitary evolution does
not fit in with the evolution predicted via projection postulate (see, e.g.,
Ref. 25).

For the sake of simplicity, let us consider a discrete, nondegenerate
observable $\mathcal{A}_{0}$ with eigenvalues a$_{0}$, a$_{1}$, a$_{2}$, ...
and let $\mid \psi _{0}\rangle $, $\mid \psi _{1}\rangle $, $\mid \psi
_{2}\rangle $, ... be the vectors representing the statuses of a macroscopic
apparatus measuring $\mathcal{A}_{0}$ that correspond to a$_{0}$, a$_{1}$, a$%
_{2}$, ... respectively. Furthermore, let $\mid \varphi _{1}\rangle $, $\mid
\varphi _{2}\rangle $, ... be vectors in the Hilbert space of the
microscopic system x such that, if x is in the state described by $\mid
\varphi _{i}\rangle $, either it is not detected (outcome a$_{0}$) or the
outcome a$_{i}$ is obtained. Whenever the initial state of the whole system
before the measurement is represented by the product vector $\mid \chi
_{i}^{(i)}\rangle $ = $\mid \varphi _{i}\rangle \mid \psi _{0}\rangle $, the
final state at the end of the measurement is represented by the vector

\begin{equation}
\mid \chi _{i}^{(f)}\rangle=t_{i}\mid \varphi _{i}\rangle \mid
\psi _{i}\rangle +t_{i}^{^{\prime }}\mid \varphi _{i}^{^{\prime }}\rangle
\mid \psi _{0}\rangle ,
\end{equation}

\noindent where $\mid t_{i}\mid ^{2}$ and $\mid t_{i}^{^{\prime }}\mid
^{2}=1-\mid t_{i}\mid ^{2}$ are the probabilities of the a$_{i}$ and the a%
$_{0}$ outcomes, respectively, and $\mid \varphi _{i}^{^{\prime }}\rangle $
is an unknown final state of the microscopic system. If one assumes that the
whole system undergoes unitary evolution and that its initial state is
represented by the product vector $\mid \chi ^{(i)}\rangle $ = $%
(\sum_{i}c_{i}\mid \varphi _{i}\rangle )\mid \psi _{0}\rangle $, the final
state is represented by the vector

\begin{equation}
\mid \chi ^{(f)}\rangle=\sum_{i}c_{i}t_{i}\mid \varphi
_{i}\rangle \mid \psi _{i}\rangle +\sum_{i}c_{i}t_{i}^{^{\prime }}\mid
\varphi _{i}^{^{\prime }}\rangle \mid \psi _{0}\rangle .
\end{equation}

\noindent If one requires consistency with the standard picture, the state S$%
_{f}$ represented (up to a normalization constant) by the vector $%
\sum_{i}c_{i}t_{i}\mid \varphi _{i}\rangle \mid \psi _{i}\rangle $ should
coincide with the state predicted by standard QM. The above description
shows that S$_{f}$ is generally different from the final state of \textit{all%
} samples at the end of the measurement, and refers to the samples of the
whole system in which the a$_{0}$ outcome does not occur, that are selected
by the observer through direct inspection of the apparatus (which does not
introduce contradictions because of the epistemic conception of states
pointed out in Sec. 5; note that the selection performed by the observer
plays the role of a second measurement on the whole system, without
implying, however, any problematic objectification induced by the conscience
of the observer himself).

Let us remind now that standard QM shows that one cannot distinguish an
entangled state from the mixture corresponding to it (via biorthogonal
decomposition$^{8}$) by simply considering probabilities of properties of
the component subsystems separately.$^{(26)}$ But if one considers in
standard QM the entangled state S$_{f}$ produced by unitary evolution at the
end of a measurement, one may maintain that S$_{f}$ can be distinguished
from the corresponding mixture M$_{S_{f}}$ predicted via projection
postulate by means of repeated measurements of the support F$_{S_{f}}$ of S$%
_{f}$ on set of samples of the whole system consisting of the microscopic
system plus the macroscopic measuring apparatus (see end of Sec. 2). This is
however impossible in practice. Alternatively, one could measure a number of
correlation properties by making measurements of first order properties on
the two subsystems separately (see again Sec. 2; it may also occur that
different correlation properties require measuring different
noncommeasurable observables on the macroscopic apparatus). But also this
procedure is practically impossible. Hence, it is reasonable to maintain
within the SR model that the description of the samples in which the a$_{0}$
outcome does not occur by means of M$_{S_{f}}$ is an approximation, which is
equivalent FAPP (for all practical purposes) to the description provided by S%
$_{f}$. Thus, because of objectivity, stochastic evolution can be seen as an
approximate law, valid for a special class of measuring devices (performing
ideal measurements) and in accessible physical situations, which avoids the
complications of a complete quantum description of the measurement process
and is hardly distinguishable from the correct theoretical law.

The problems raised by the projection postulate are thus greatly
undramatized in our perspective. Moreover, this postulate provides a correct
description of the microscopic system after a measurement
in standard QM (see (ii) above), \textit{hence also within the SR model if
one considers only physical objects that are actually detected}. We
therefore close this section pointing out some consequences of it in the
framework of the SR model and showing that these consequences are consistent
with the general remarks on the measurement process made in Sec. 5.

To begin with, let us observe that one can introduce a physically meaningful
relation of \textit{consistency} on the set of all pure states of a physical
system in the SR model by saying that the pure states S$_{1}$ and S$_{2}$
are consistent iff the vectors $\mid \varphi _{1}\rangle $ and $\mid \varphi
_{2}\rangle $ that represent them, respectively, are not orthogonal. Indeed,
if S$_{1}$ and S$_{2}$ are consistent, there is no property that is
certainly possessed by a physical object x if x is in the state S$_{1}$ and
certainly not possessed if x is in the state S$_{2}$. On the contrary, such
a property exists if S$_{1}$ and S$_{2}$ are not consistent. Thus, S$_{1}$
and S$_{2}$ are consistent iff x can possess the same properties in the
state S$_{1}$ and in the state S$_{2}$ (of course, we cannot know whether
this actually occurs, since the knowledge of the state of a physical object
provides only incomplete information about its properties, see Sec. 5).

Then, the description of measurements provided by the projection postulate
has the following features in the SR model (see also Ref. 3).

(i) The final pure state S$_{j}$ of a physical object x after a measurement
of an observable $\mathcal{A}_{0}$ that yields the (possibly degenerate)
eigenvalue a$_{j}$ is consistent with the pure state S of x before the
measurement. Indeed, S and S$_{j}$ are represented by non-orthogonal
vectors. This implies that x may possess the same properties before and
after the measurement. Hence, \textit{the measurement produces a change of
our information about x, but it does not necessarily imply a change of the
properties of x} (though a change of the set of properties possessed by x 
\textit{may} actually occur because of the interaction of the physical
object with the measuring apparatus). We thus get, by using the projection
postulate, the same result that we have obtained by means of general
arguments in Sec. 5 (which introduces, as we have already observed, a
viewpoint that is deeply different from the viewpoint adopted by the
standard quantum measurement theory).

Note that if S$_{j}$ and S$_{k}$ are the pure states predicted by the
projection postulate after distinct measurements on a physical object x in
the state S that yield the eigenvalues a$_{j}$ and a$_{k}$ of the observable 
$\mathcal{A}_{0}$, respectively, S$_{j}$ and S$_{k}$ are not consistent,
since the vectors representing them are orthogonal. This matches with the
fact that the properties F$_{S_{j}}$ = ``having value a$_{j}$ of $\mathcal{A}%
_{0}$'' and F$_{S_{K}}$ = ``having value a$_{k}$ of $\mathcal{A}_{0}$'',
that are the supports of S$_{j}$ and S$_{k}$, respectively, are mutually
exclusive.

(ii) Let S$_{a}$ and S$_{b}$ be the pure states predicted by the projection
postulate after (ideal) measurements on a physical object x in the state S
that yield the eigenvalues a and b of the observables $\mathcal{A}_{0}$ and $%
\mathcal{B}_{0}$, respectively. Then, S$_{a}$ and S$_{b}$ can be consistent
even when the properties F$_{a}$ = ``having value a of $\mathcal{A}_{0}$''
and F$_{b}$ = ``having value b of $\mathcal{B}_{0}$'', that are the supports
of S$_{a}$ and S$_{b}$, respectively, are noncommeasurable. This occurs
whenever the vectors $\mid \varphi _{a}\rangle $ and $\mid \varphi
_{b}\rangle $ that represent S$_{a}$ and S$_{b}$, respectively, are not
orthogonal, and intuitively means that x may possess both properties F$_{a}$
and F$_{b}$ conjointly, even if one cannot generally know whether this
occurs by means of a conjoint measurement of them (yet, one can attain this
knowledge in some cases by means of a prediction followed by a measurement,
as we have seen in Sec. 5). Again, this feature distinguishes the viewpoint
provided by the SR model from the standard interpretation, according to
which F$_{a}$ and F$_{b}$ can never be simultaneously real for the physical
object x.

$\,$

\section{\normalsize{AN INTUITIVE PICTURE FOR THE SR MODEL}}

Our treatment of the locality and measurement problems in the previous
sections has been carried out by referring to the SR model in which
macroscopic properties only are considered, so that strict operational
requirements are fulfilled (though the SR model does not adopt a
verificationist attitude, see Ref. 1). Therefore, our perspective does not
provide an intuitive picture of what is going on at the microscopic level.
But if one accepts introducing microscopic properties of physical objects
such a picture becomes possible and it has been recently propounded by one
of us as an autonomous model$^{(27)}$, based on the \textit{extended SR model%
} expounded in Ref. 1 but bringing in it some important corrections. The new
model provides a sample of objective interpretation of QM, and all relevant
features of the SR model hold in it (in particular, MGP), so that it can be
regarded from our present viewpoint as a set-theoretical proof of the
consistency of the SR model. For the sake of completeness we therefore
report the essentials of it here.

To begin with, we accept the correspondence of microscopic and macroscopic
properties established in the framework of the extended SR model. To be
precise, we assume that every microscopic physical system is characterized
by a set $\mathcal{E}$ of microscopic physical properties (which play the
role of theoretical entities), and that every sample of the system (physical
object) either possesses or does not possess each property in $\mathcal{E}$.
Moreover, every microscopic property \textit{f} in $\mathcal{E}$ corresponds
to a macroscopic property F = ($\mathcal{A}_{0}$,$\Delta $), where $\mathcal{%
A}_{0}$ is an observable and $\Delta $ a Borel set on the real line such
that a$_{0}\notin \Delta $, hence is represented by the same projection
operator that represents ($\mathcal{A}_{0}$,$\Delta $). Whenever a physical
object x is prepared by a given preparing device $\pi $ (for the sake of
simplicity we assume here that the equivalence class of $\pi $ is a pure
state S, see Ref. 1, Sec. 2) and $\mathcal{A}_{0}$ is measured by means of a
suitable apparatus, the set of microscopic properties possessed by x
produces a probability (which is either 0 or 1 if the model is \textit{%
deterministic}) that the apparatus does not react, so that the outcome a$%
_{0} $ may be obtained. In this case, a nonaccessible physical situation
occurs, and we cannot get any explicit information about the microscopic
physical properties possessed by x. In particular, we cannot assert that
they are related as the projection operators representing them are related
by the laws of standard QM, which is consistent with MGP. If, on the
contrary, the apparatus reacts, an outcome different from a$_{0}$, say a, is
obtained, and we are informed that x possesses all microscopic properties
associated with macroscopic properties of the form F = ($\mathcal{A}_{0}$,$%
\Delta $), where $\Delta $ is a Borel set such that a$_{0}\notin \Delta $
and a $\in \Delta $ (for the sake of brevity, we also say that x \textit{%
possesses} all macroscopic properties as F in this case). Then, whenever a
law of standard QM is considered, both accessible and nonaccessible physical
situations may occur, and only in the former situations we can assert that
the microscopic properties of x are related as the projection operators
representing them in the given law.

Let us come now to our intuitive picture. Whenever the preparing device $\pi 
$ is activated repeatedly, a (finite) set $\mathcal{S}$ of physical objects
in the state S is prepared. Let us partition $\mathcal{S}$ into subsets $%
\mathcal{S}^{(1)}$, $\mathcal{S}^{(2)}$, ..., $\mathcal{S}^{(n)}$, such that
in each subset all objects possess the same \textit{microscopic} properties,
and assume that a measurement of an observable $\mathcal{A}_{0}$ is done on
every object. Furthermore, let us introduce the following symbols.

N: number of physical objects in $\mathcal{S}$.

N$_{0}$: number of physical objects in $\mathcal{S}$ that are not detected.

N$^{(i)}$: number of physical objects in $\mathcal{S}^{(i)}$.

N$_{0}^{(i)}$: number of physical objects in $\mathcal{S}^{(i)}$ that are
not detected.

N$_{F}^{(i)}$: number of physical objects in $\mathcal{S}^{(i)}$ that
possess the macroscopic property F = ($\mathcal{A}_{0}$,$\Delta $)
corresponding to the microscopic property \textit{f}.

It follows from our above interpretation that the number N$_{F}^{(i)}$
either coincides with N$^{(i)}-$N$_{0}^{(i)}$ or with 0. The former case
occurs whenever \textit{f} is possessed by the objects in $\mathcal{S}^{(i)}$%
, since all objects that are detected then yield outcome in $\Delta $. The
latter case occurs whenever \textit{f} is not possessed by the objects in $%
\mathcal{S}^{(i)}$, since all objects that are detected then yield outcome
different from a$_{0}$ but outside $\Delta $. In both cases one generally
gets N$^{(i)}-$N$_{0}^{(i)}$ $\neq $ 0 (even if N$^{(i)}-$N$_{0}^{(i)}$ = 0
may also occur, in particular in a deterministic model), so that the
following equation holds:

\begin{equation}
\frac{N_{F}^{(i)}}{N^{(i)}}=\frac{N^{(i)}-N_{0}^{(i)}}{N^{(i)}}\frac{%
N_{F}^{(i)}}{N^{(i)}-N_{0}^{(i)}} .
\end{equation}

The term on the left in Eq. (3) represents the frequency of objects
possessing the property F in $\mathcal{S}^{(i)}$, the first term on the
right the frequency of objects in $\mathcal{S}^{(i)}$ that are detected, the
second term (which either is 1 or 0) the frequency of objects that possess
the property F in the subset of all objects in $\mathcal{S}^{(i)}$ that are
detected.

The frequency of objects in $\mathcal{S}$ that possess the property F is
given by

\begin{equation}
\frac{1}{N}\sum_{i}N_{F}^{(i)}=\frac{N-N_{0}}{N}(\sum_{i}\frac{%
N_{F}^{(i)}}{N-N_{0}}) .
\end{equation}
Let us assume now that all frequencies converge in the large number limit,
so that they can be substituted by probabilities, and that these
probabilities do not depend on the choice of the preparation $\pi $ in S
(which is consistent with the definition of states in Ref. 1, Sec. 2).
Hence, if one considers the large number limit of Eq. (3), one gets

\begin{equation}
\mathcal{P}_{S}^{(i)t}(F)=\mathcal{P}^{(i)d}(F)\mathcal{%
P}_{S}^{(i)}(F) ,
\end{equation}

\noindent where $\mathcal{P}_{S}^{(i)t}(F)$ is interpreted as the overall
probability that a physical object x possessing the microscopic properties
that characterize $\mathcal{S}^{(i)}$ also possess the property F, $\mathcal{%
P}_{S}^{(i)d}(F)$ as the probability that x be detected when F is measured
on it, $\mathcal{P}_{S}^{(i)}(F)$ (which either is 0 or 1) as the
probability that x possess the property F when detected. Moreover, if one
considers the large number limit of Eq. (4) and reminds the interpretation
of quantum probabilities in Sec. 4, (1), it is reasonable to assume that the
second term on the right converges to the standard quantum probability $%
\mathcal{P}_{S}(F)$ that a physical object in the state S possess the
property F, so that one gets

\begin{equation}
\mathcal{P}_{S}^{t}(F)=\mathcal{P}^{d}(F)\mathcal{P}%
_{S}(F) ,
\end{equation}

\noindent where $\mathcal{P}_{S}^{t}(F)$ is interpreted as the overall
probability that a physical object x in a state S possess the property F and 
$\mathcal{P}_{S}^{d}(F)$ as the probability that x be detected when F is
measured on it. Thus, one can maintain that a broader theory embodying QM
can be conceived, according to which the standard quantum probability $%
\mathcal{P}_{S}(F)$ is considered as a \textit{conditional} rather than an 
\textit{absolute} probability. Of course, Eq. (6) is also compatible with
a model in which $\mathcal{P}^{d}(F)$ is interpreted as the
efficiency of a non-ideal measuring apparatus. Yet, our picture predicts
that $\mathcal{P}^{d}(F)$ may be less than 1 also in the case
of an ideal apparatus. Indeed, every $\pi $ in S prepares objects which do
not possess the same microscopic properties, and some objects may possess
sets of properties that make the detection of them by any apparatus
measuring F possible but not certain, or even impossible.

To close up, let us note that the intuitive picture provided above
introduces a substantial correction in the extended SR model that inspires
it, since it substitutes Eq. (5) to the equation $\mathcal{P}_{\mathcal{A}%
_{0}S}(F)$ = $\mathcal{P}_{\mathcal{A}_{0}}(F,G,H,...)\mathcal{P}_{S}(F)$
that appears in this model. Bearing in mind its definition, the probability $%
\mathcal{P}_{\mathcal{A}_{0}S}(F)$ can be obtained as the large number limit
of the term on the left in Eq. (3) (which shows that the suffix \textit{S}
in it is rather misleading). Analogously, the probability $\mathcal{P}_{%
\mathcal{A}_{0}}(F,G,H,...)$ can be obtained as the large number limit of
the first term on the right in the same equation. However, the second term
on the right in Eq. (3) does not converge to the quantum probability $%
\mathcal{P}_{S}(F)$, hence the above equation of the extended SR model is
not correct in our new picture. We observe, however, that also in this
picture the \textit{fair sampling assumption} does not hold, which is
important for theoretical reasons (see Ref. 1, Sec. 5).\medskip
\vspace{5cm}

\noindent
\textbf{\normalsize{AKNOWLEDGEMENTS}\smallskip }

We wish to thank Prof. Carlo Dalla Pozza, Prof. Arcangelo Rossi and Prof.
Luigi Solombrino for valuable help and suggestions.

\section*{\normalsize{REFERENCES}}

\begin{enumerate}
\item
C. Garola, ``A simple model for an objective interpretation of
quantum mechanics,'' \textit{Found. Phys. }\textbf{32, }1597-1615 (2002).

\item
C. Garola, ``Classical foundations of quantum logic,'' \textit{%
Internat. J. Theoret. Phys.} \textbf{30,} 1-52 (1991).

\item
C. Garola and L. Solombrino, ``The theoretical apparatus of
Semantic Realism: a new language for classical and quantum physics,'' 
\textit{Found. Phys. }\textbf{26, }1121-1164 (1996).

\item
C. Garola and L. Solombrino, ``Semantic Realism versus
EPR-like paradoxes: the Furry, Bohm-Aharonov and Bell paradoxes,'' \textit{%
Found. Phys. }\textbf{26, }1329-1356 (1996).

\item
C. Garola, ``Objectivity versus nonobjectivity in quantum
mechanics,'' \textit{Found. Phys}. \textbf{30}, 1539-1565 (2000).

\item
E. G. Beltrametti and G. Cassinelli, \textit{The logic of
quantum mechanics} (Addison-Wesley, Reading, MA, 1981).

\item
F. Selleri, ``History of the Einstein-Podolski-Rosen
paradox,'' in \textit{Quantum Mechanics Versus Local Realism}, F. Selleri,
ed. (Plenum, New York, 1988).

\item
A. Einstein, B. Podolski, and N. Rosen, ``Can
quantum-mechanical description of physical reality be considered
complete?,'' \textit{Phys. Rev}. \textbf{47}, 777-780 (1935).

\item
D. Bohm, \textit{Quantum Theory} (Prentice-Hall, Englewood
Cliffs, NJ, 1951).

\item
J. F. Clauser and M. A. Horne, ``Experimental consequences of
objective local theories,'' \textit{Phys. Rev. D }\textbf{10}, 526-535
(1974).

\item
J. S. Bell, ``On the Einstein Podolski Rosen Paradox,'' 
\textit{Physics} \textbf{1}, 195-200 (1964).

\item
M. Redhead, \textit{Incompleteness, Nonlocality and Realism}
(Clarendon, Oxford, 1987).

\item
E. P. Wigner, ``On hidden variables and quantum mechanical
probabilities,'' \textit{Amer. J. Phys}. \textbf{38}, 1005-1009 (1970).

\item
J. J. Sakurai, \textit{Modern Quantum Mechanics} (Benjamin,
Reading, 1985).

\item
J. F. Clauser, M. A. Horne, A. Shimony, and R. A. Holt,
``Proposed experiment to test local hidden variables theories,'' \textit{%
Phys. Rev. Lett}. \textbf{23}, 880-884 (1969).

\item
D. M. Greenberger, M. A. Horne, A. Shimony, and A.
Zeilinger, ``Bell's Theorem without Inequalities,'' \textit{Amer. J. Phys.} 
\textbf{58}, 1131-1143 (1990).

\item
N. D. Mermin, ``Hidden variables and the two theorems of
John Bell,'' \textit{Rev. Modern Phys.} \textbf{65,} 803-815 (1993).

\item
C. Garola, ``Reconciling local realism and quantum physics: a
critique to Bell,'' \textit{Theoret. Mathem. Phys.} \textbf{99}, 285-291
(1994).

\item
C. Garola, ``Criticizing Bell: local realism and quantum
physics reconciled,'' \textit{Internat. J. Theoret. Phys.} \textbf{34},
253-263 (1995).

\item
C. Garola, ``Questioning nonlocality: an operational critique
to Bell's theorem,'' in \textit{The Foundations of Quantum Mechanics.
Historical Analysis and Open Questions}, C. Garola and A. Rossi, eds.
(Kluwer, Dordrecht, 1995).

\item
C. Garola, ``Essay review: waves, information and foundations
of physics,'' \textit{Studies in History and Philosophy of Modern Physics} 
\textbf{33}, 101-116 (2002).

\item
S. Kochen and E. P. Specker, ``The problem of hidden
variables in quantum mechanics,'' in \textit{The Logico-Algebraic Approach
to Quantum Mechanics I}, C. A. Hooker, ed. (Reidel, Dordrecht, 1975).

\item
A. Aspect, ``Experimental tests of Bell's inequalities with
correlated photons,'' in \textit{Waves, Information and Foundations of
Physics}, R. Pratesi and L. Ronchi, eds. (Editrice Compositori, Bologna,
1998).

\item
C. Garola, ``Semantic realism: a new philosophy for quantum
physics,'' \textit{Internat. J. Theoret. Phys}. \textbf{38}, 3241-3252
(1999).

\item
D. Z. Albert, \textit{Quantum Mechanics and Experience}
(Harvard University Press, Cambridge, MA, 1992).

\item
B. D'Espagnat, \textit{Conceptual Foundations of Quantum
Mechanics} (Benjamin, Reading, MA, 1976).

\item
C. Garola, ``Embedding quantum mechanics into an objective framework,'' 
\textit{Found. Phys. Lett.} \textbf{16}, 605-612 (2003).
\end{enumerate}
\end{document}